\documentclass[twocolumn,prb,amsmath,amssymb,floatfix]{revtex4-1}

\usepackage{amsmath}%
\usepackage{amsfonts}%
\usepackage{amssymb}%
\usepackage{graphicx}% Include figure files
\usepackage{dcolumn}% Align table columns on decimal point
\usepackage{bm}% bold math
\usepackage{mathrsfs}

\DeclareMathOperator{\erfc}{erfc}

\begin{document}

\title{Long Range Ballistic Motion and Coherent Flow of Long Lifetime Polaritons}

\author{Mark Steger}
\email{mds71@pitt.edu.}
\author{Gangqiang Liu}
\author{Bryan Nelsen}
\author{Chitra Gautham}
\author{David W. Snoke}
 
\affiliation{Department of Physics and Astronomy, University of Pittsburgh, 3941 O'Hara Street, Pittsburgh, Pennsylvania 15260, USA}

\author{Ryan Balili}
 
\affiliation{NanoPhotonics Centre, Cavendish Laboratory, University of Cambridge, J. J. Thomson Avenue, Cambridge CB3 0HE, United Kingdom}

\author{Loren Pfeiffer}
\author{Ken West}

\affiliation{Department of Electrical Engineering, Princeton University, Princeton, New Jersey 08544, USA}

\date{\today}

\begin{abstract}
Exciton-polaritons can be created in semiconductor microcavities. These quasiparticles act as weakly interacting bosons with very light mass, of the order of $10^{-4}$ times the vacuum electron mass. Many experiments have shown effects which can be viewed as due to a Bose-Einstein condensate, or quasicondensate, of these particles. The lifetime of the particles in most of those experiments has been of the order of a few picoseconds, leading to significant nonequilibrium effects. By increasing the cavity quality, we have made new samples with longer polariton lifetimes.  With a photon lifetime on the order of 100-200 ps, polaritons in these new structures can not only come closer to reaching true thermal equilibrium, a desired feature for many researchers working in this field, but they can also travel much longer distances.  We observe the polaritons to ballistically travel on the order of one millimeter, and at higher densities we see transport of a coherent condensate, or quasicondensate, over comparable distances.  In this paper we report a quantitative analysis of the flow of the polaritons both in a low-density, classical regime, and in the coherent regime at higher density. Our analysis gives us a measure of the intrinsic lifetime for photon decay from the microcavity and a measure of the strength of interactions of the polaritons. 
\end{abstract}

\maketitle

\section{Introduction}

When a photon in a cavity is energetically resonant or nearly resonant with an exciton state, a mixed state known as an exciton-polariton arises\cite{Deng2010, Kavokin2007, Snoke2010}.  This new quasiparticle has a light mass, of the order of $10^{-4}$ times the vacuum electron mass, which it inherits from the cavity photon dispersion, but has particle-particle interactions comparable in magnitude to exciton-exciton interactions, which are much stronger than standard photon-photon nonlinear interactions \cite{Snoke2012}.  Exciton-polaritons provide an appealing system for the study of bosonic statistics as well as a platform for quantum or optical circuit components.  The very light mass of the polaritons means that bosonic effects can be relevant at much higher temperatures than those required for atomic condensates, typically tens of Kelvin up to room temperature.  The particles can be created with an incoherent or coherent source and can be subjected to optically or mechanically generated potential barriers or traps controlled by the researcher, allowing for great flexibility \cite{Balili2006a,Tosi2012, Hayat2012}.  Patterning of the samples also allows for arbitrary lattice structures of the polaritons \cite{Masumoto2012}, and nanostructures can be etched into the sample to make waveguides and circuits to control the propagation and confinement of polaritons \cite{Bloch1997, Bloch1998, Wertz2010}.

In a typical structure, such as used for these studies, a GaAs-based microcavity is designed to include GaAs quantum wells (QWs) located at the antinodes of a planar optical cavity with end mirrors which are distributed Bragg reflectors (DBRs) made from Al$_x$Ga$_{1-x}$As and AlAs layers.  The strong coupling of the exciton to the photon through the exciton's radiative dipole matrix element leads to the formation of new states called the upper and lower polaritons (UP and LP).  Our sample is a $3 \lambda /2$ microcavity containing 4 GaAs/AlAs QWs at each of the 3 antinodes. The DBRs confining the cavity are made of alternating planar $\lambda /4$ layers of Al$_{0.2}$Ga$_{0.8}$As/AlAs. This sample is similar to one used in previous work\cite{Balili2007}, but the number of layers in both the front and back DBRs were doubled, effectively increasing the designed Q-factor by more than two orders of magnitude and the designed photon lifetime from 2 ps to 400 ps.  This is the same sample as studied in Ref.~\onlinecite{Nelsen2013PRXSubmission}.

For polariton experiments, the pumping method has important implications.  Polaritons can be pumped resonantly with a laser matched in energy and angle with the polariton dispersion relation, or they can be pumped non-resonantly at much higher energy where the stop band of the DBRs becomes transparent.  Resonant excitation of polaritons can potentially seed a condensate to form in user-selected states or impart initial coherence to a population, since the polaritons generated will initially have the same coherence characteristics as the exciting photons \cite{Hayat2013}.  However, non-resonant excitation can serve as a more concrete demonstration of Bose-Einstein condensation and related effects, since the initially generated carriers lose the coherence of the pump source while relaxing to a thermal quasi-equilibrium in the polariton states.  If coherence is observed to increase with increasing density or a macroscopic occupation of one state forms out of the thermal background, then this suggests a spontaneous symmetry breaking which is not present in resonant excitation experiments.

Additionally, the hot carriers and excitons generated during non-resonant excitation can lead to other interesting physics.  The repulsive exciton-exciton interaction increases the exciton energy, and phase space filling leads to a reduction of the exciton-photon coupling.  Both of these features serve to increase the energy of the LP, so that at moderate and high pump power the LP sees a modified energy potential at the excitation spot.  This optically generated barrier has been used to modulate the polariton profile and dynamics \cite{Richard2005, Richard2005b, Tosi2012, Christmann2012, Wen2013Submission}.

Another crucial choice in the experiments is the detuning, that is, the energy difference between the bare photon and exciton.  The detuning determines the relative fraction of photon and exciton in the polaritons. In typical GaAs and other semiconductor wafers grown with molecular beam epitaxy, the thickness of the layers varies across the wafer. Since the photon energy and the exciton energy have different dependences on the layer thickness, this allows a design in which the exciton and photon energies cross at some place on the wafer. On one side of this resonant region where the energies cross, the photon energy is lower than the exciton energy, and the LP will have a mostly photonic character. On the other side, the photon energy is higher than the exciton energy, and the character of the LP is mostly excitonic. On the excitonic side, the interactions are stronger, and the mass is heavier leading to greater thermalization \cite{Deng2006,Wen2013Submission} but shorter distances for transport of the polaritons. On the photonic side, the interactions are smaller, allowing less thermalization, but much longer transport distances. For the experiments discussed in this paper, we chose a location on the wafer where the LP was mostly photonic, allowing long distance transport. The polaritons still interact with each other and with the excitons at the generation spot, as we will show below. 

\section{Low Density: Ballistic Propagation}

\begin{figure}[h]
 \begin{center}  
  \includegraphics[width=3.35in]{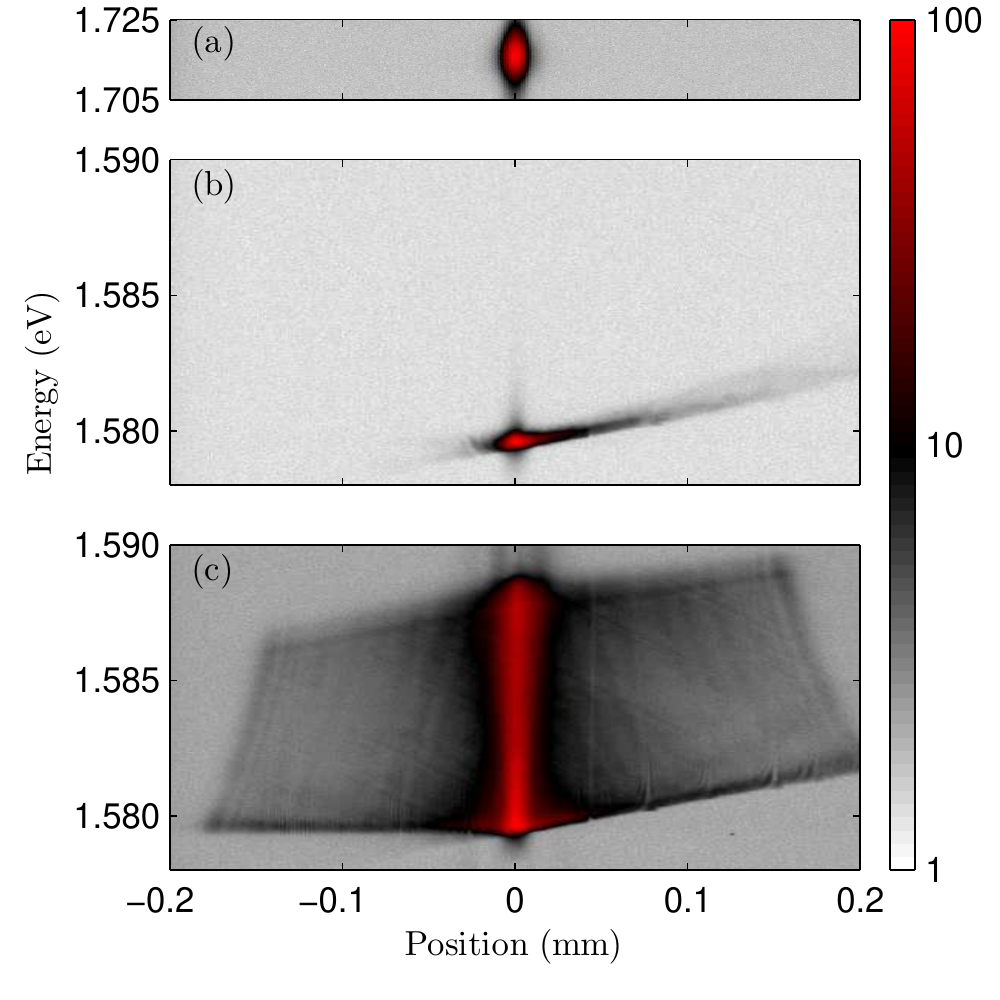}
  \caption{(color online) Intensity of the PL emitted from the LP, as a function of energy $E$ and position $x$,  recorded with an imaging spectrometer.  The intensity data are presented on a log scale to highlight motion.  These data are taken at a $k_{\|}=0$ polariton detuning of -21 meV with a pump power of 500 $\mu$W at 705 nm focused to a 15 $\mu$m diameter spot size. a) Hot-carrier luminescence seen through the reflectivity minima of the DBR stop band.  The size of this spot indicates the size of the pump spot and the exciton cloud. b) Lower polariton PL, spatially resolved but only collected near $k_{||}=0$.  The bright spot is the point of creation of the polaritons;  the PL at further distances gives the $k_{||}=0$ energy of polaritons which have moved to that point on the sample. c) The same data taken with a larger NA, i.e., a larger range of $k_{\|}$. The polaritons flow outward from the creation spot to fill all space within our field of view.  The polaritons are generated over a broad range of $k_{\|}$ at the pump spot and ballistically travel outward at constant energy.  The sharp cutoff in energy on the downhill side indicates that the polaritons do not scatter once they are spatially distant from the excitation spot.  The horizontal/angled cutoff at high energy is the accepted NA of the microscope objective.  The cutoffs at $\pm$ 0.2 mm are due to clipping in the optics and spectrometer.}
  \label{EvX_LowDensity}
 \end{center}
\end{figure}

The first observation of polariton photoluminescence (PL) in these samples was initially perplexing.  Luminescence data in Fig.~\ref{EvX_LowDensity}(b) shows polaritons on the LP branch propagating a long distance on the sample from the excitation spot. Looking only at this figure, it appears that the polaritons gain energy to travel uphill.  

If we compare this to Fig.~\ref{EvX_LowDensity}(c), however, we can make more sense of the data.  Figure~\ref{EvX_LowDensity}(b) was taken with small numerical aperture (NA), while Fig.~\ref{EvX_LowDensity}(c) was taken with large NA.  The NA matters because a polariton with wavevector $k_{\|}$ is a coupling of an exciton and a cavity photon both with the same $k_{\|}$; when the polariton decays, it emits a photon external to the cavity with the same wavevector. This gives a one-to-one mapping of the angle of the photon emission in the far field to the in-plane $k_{\|}$ of the polaritons before they decay into external photons. Therefore opening up the numerical aperture of the imaging system collects light from polaritons at higher $k_{\|}$. For the data of Fig.~ \ref{EvX_LowDensity}(b), the low NA restricted the polaritons observed to those with $k_{\|} \thicksim 0$.  We see in this figure the gradient of the $k_{\|}=0$ energy, i.e., the potential energy of the polaritons, due to the wedge in the wafer thickness discussed above. This spatial gradient of the ground state energy is the same as a force on the polaritons, since $F = -\nabla U$.

The data of \ref{EvX_LowDensity}(c) was taken with a lens system with a 0.4 NA, much larger than the NA used for Fig.~\ref{EvX_LowDensity}(b).  This larger acceptance angle corresponds to imaging polaritons with a much wider range of momenta.   Figure~\ref{EvX_LowDensity}(c) shows that there is a significant population of polaritons at $k_{\|} > 0$; The broad distribution of high-momentum polaritons exists at the point of creation due to the many random scattering processes which occur after non-resonant excitation. Some of the high-$k_{\|}$ polaritons flow uphill and eventually reach $k_{\|} =0$ where they can be observed with low NA, while others flow downhill until they exit the 0.4 NA collection angle.

One critical feature to notice in these data is the sharp minimum energy cutoff on the left side of Fig.~\ref{EvX_LowDensity}(c).  The polaritons at the excitation spot partially thermalize according to the relaxation dynamics of hot carriers and excitons \cite{Hartwell2008,Hartwell2010}.  Upon reaching polariton states with very light mass and low scattering rates, the polaritons are able to travel ballistically.  This explains the minimum energy observed on the right---polaritons are streaming ballistically away from the excitation spot after initially scattering into LP states.  The polaritons flowing downhill immediately leave the high density excitation region and never scatter to lower energy. The polaritons flowing uphill stream until they hit a point on the sample where the $k_{\|} = 0$ state has the same energy, at which point they can no longer flow to the right, and are reflected back to the left. 

\begin{figure}[h]
 \begin{center}  
  \includegraphics[width=3.35in]{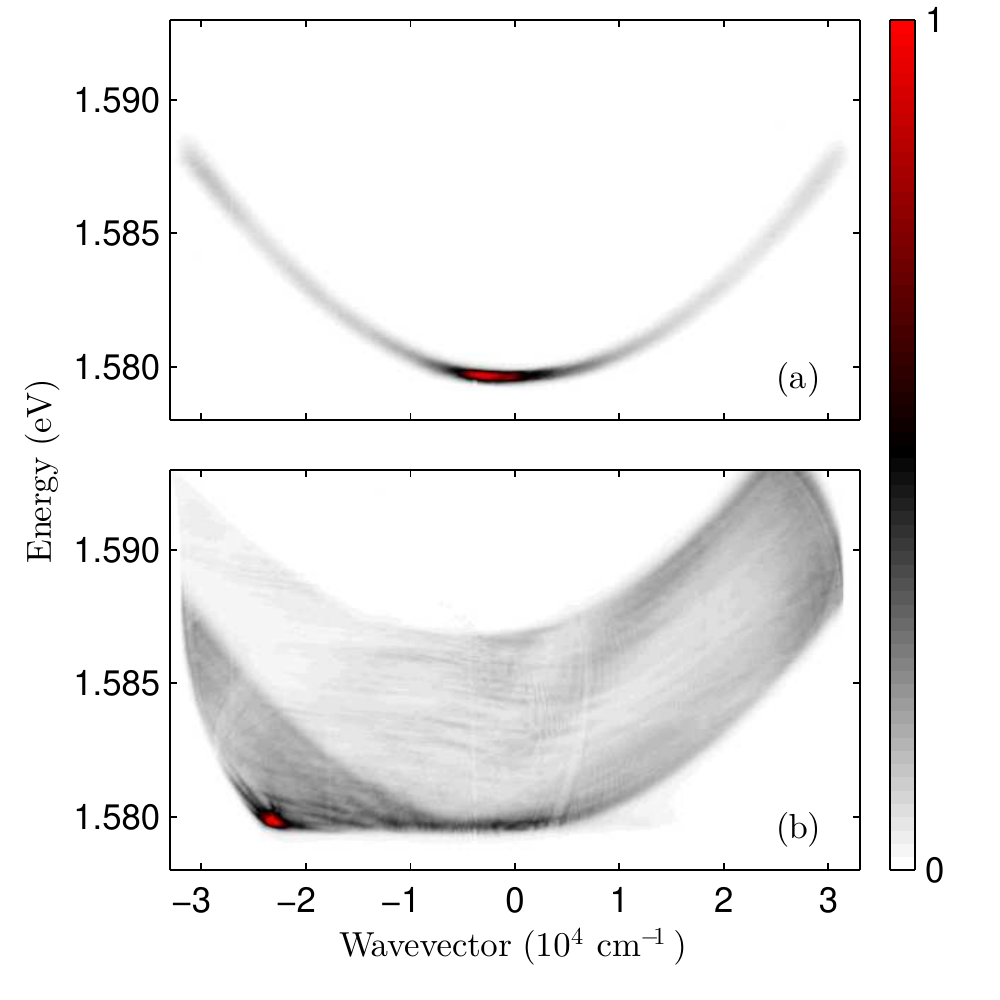}
  \caption{(color online) Intensity of the PL emitted from the LP, as a function of energy $E$ and in-plane momentum $k_{\|}$, recorded using an imaging spectrometer focused on the far-field emission (Fourier plane).  This data was taken under the same pumping conditions as Fig.~\ref{EvX_LowDensity}. a) Spatial filtering is applied to an intermediate image to isolate the dispersion relation of the LP at the pump spot.  Due to non-resonant excitation, polaritons are observed filling the momentum states. b) With no spatial filtering, the excitation spot polaritons are smeared in the downhill ($-k_{\|}$) direction.  Polaritons at $k_{\|}=0$ correspond to the polaritons observed in Fig.~\ref{EvX_LowDensity}(b).  Again we observe an energy minimum coinciding with the vertex of the pump spot dispersion curve, as the polaritons scatter very little after leaving the creation region.}
  \label{EvK_LowDensity}
 \end{center}
\end{figure}

Because of the one-to-one mapping of polariton momentum to photon emission angle in the decay process, we can image the far-field PL to directly resolve the momentum space distribution of the polaritons, just as we image the near field to observe the real-space profile.  Normally, the $k$-space image integrates over the entire real space observed, so we must use spatial filtering to measure the dispersion relation from a single point of a spatially extended distribution. In Fig.~\ref{EvK_LowDensity}(a) we present the far-field PL of the emission spot after spatial filtering, which was accomplished using a pinhole in a secondary real image plane.  The spatial filter selects a region on the sample of approximately 40 $\mu$m diameter, which is slightly larger than the pump spot.  The PL profile at this spot indicates the initial population before propagating away.  Figure \ref{EvK_LowDensity}(b) shows the same data without the spatial filter.

There are several features of Fig.~\ref{EvK_LowDensity}(b) which are complementary to the real-space data of Fig.~\ref{EvX_LowDensity}(c).  The polaritons initially at $+k_{\|}$ move uphill at constant energy while losing momentum, i.e., shifting to lower $k_{\|}$.  The polaritons at $-k_{\|}$ flow downhill at constant energy and gain momentum in that direction, eventually leaving the numerical aperture of our microscope objective.  There is again a clear cutoff in energy at the vertex of the excitation spot momentum dispersion parabola.  The polaritons starting at $k_{\|}=0$ are the lowest energy polaritons possible at the pump spot where the density is high enough to scatter.  These polaritons stream downhill ballistically, giving rise to this energy minimum. 

%Figure showing E-vs-k.  1 spatialy filtered, the other no filtering

\section{Time-resolved Propagation}

To verify that the extended polariton cloud is propagating from the point of excitation, we used a Hamamatsu streak camera to time-resolve the spatial arrival of polaritons at various points on the wafer following a pump pulse with picosecond duration.  Due to the many scattering processes following the non-resonant generation, all the temporal dynamics of the cool polaritons are broadened and delayed relative to the excitation pulse. Nevertheless, by measuring the arrival times of the polaritons moving uphill, we see clearly that there is a time delay for the propagation of the polaritons as they travel across the wafer. As discussed below, this time delay is consistent with the theory for the time of flight across the sample, using the known polariton dispersion.

\begin{figure}[h]
 \begin{center}  
  \includegraphics[width=3.35in]{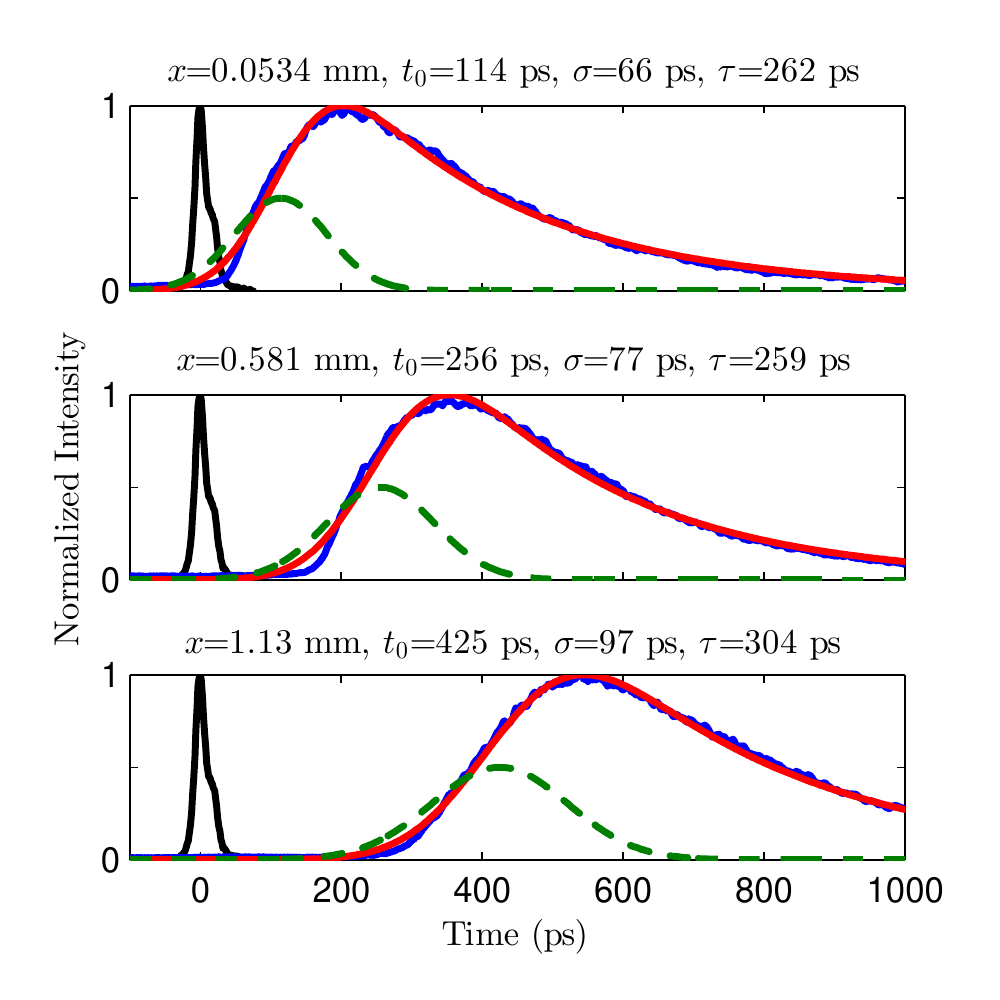}
  \caption{(color online) Time-resolved $k_{\|} \simeq 0$ PL from the lower polartion at three sample distances from the pump spot.  These data were collected following a 2 ps pump laser with wavelength of 725 nm focused to a 50 $\mu$m diameter pump spot where the $k_{\|}=0$ polariton detuning was -15 meV. Blue lines are intensity data of photoluminescence from propagating polaritons.  Each frame is taken at a different distance from the pump spot.  Black lines indicate the emission of the hot carriers above the stop band which occurs very soon after the picosecond pump.  Red lines are the Gaussian-exponential decay convolution fits to the data with the parameters given above each frame.  $t_0$ is the central time of the Gaussian following the hot PL, $\sigma$ is the standard deviation, and $\tau$ is the exponential decay time.  Note that $t_0$ is an indicator of the travel time--we know that this value must include both the time of flight as well as the time to cool down from hot carriers to the lower polariton.  As an aid to the reader, the unconvolved Gaussian is presented as the dashed green line to see how the $t_0$ parameter compares to the peak of the intensity data.  The convolution with a decay pushes the peak of the fit to significantly later time than the Gaussian fit alone.}
  \label{TimeProfiles}
 \end{center}
\end{figure}

Figure~\ref{TimeProfiles} shows the time-resolved PL for the polaritons (solid blue line) for different distances $x$ from the generation spot, following the hot PL emission (black line), which indicates the duration of the pump laser pulse. The polaritons were generated non-resonantly on the photonic side of the wafer with a 2-ps pulsed Ti:Sapphire laser, and $k_{\|}\sim 0$ emission from individual spatial points was spectrally and temporally resolved.  The polariton PL is fit with a Gaussian convolved with an exponential decay as shown with the solid red line.  The details of this fit are discussed in the Appendix.  

%Figure showing I-vs-t.

The convolution of a Gaussian and decay is an empirical fit which is sufficient for assigning an arrival time to the polaritons.  Additionally, this convolution can be written in a closed form, which makes it computationally convenient to fit the data.  Aside from background and overall amplitude, the fit has three parameters: the arrival time $t_0$, the Gaussian broadening $\sigma$, and the decay time $\tau$.  We interpret the arrival time as the sum of two major contributions: 1) first, the hot excitations cool down to fill the polariton states at the pump spot.  This cool-down time depends on the phonon emission rates. 2) The remainder of the arrival time is due to the actual time of flight (TOF) of the ballistic polaritons to reach a point on the sample where their momentum has slowed to $k_{\|}\simeq 0$, where they are observed. The decay time $\tau$ cannot simply be interpreted as the lifetime of the polaritons, since the dynamics of the hot carriers fills these states over a finite time.  For example, if the time to cool down into polariton states is comparable to or longer than the lifetime of the polaritons, then the decay time will measure the lifetime of this excited population rather than that of the polaritons.  

The green dashed line in each case of Fig.~\ref{TimeProfiles} is the Gaussian portion of the convolution. As seen in this figure, the peak of each PL curve is not at the fitted $t_0$ value, which is located at the peak of this pure Gaussian, but is shifted to a later time by the convolution with an exponential decay. 
%Figure showing x-vs-TOF.
\begin{figure}[h]
 \begin{center}  
  \includegraphics[width=3.35in]{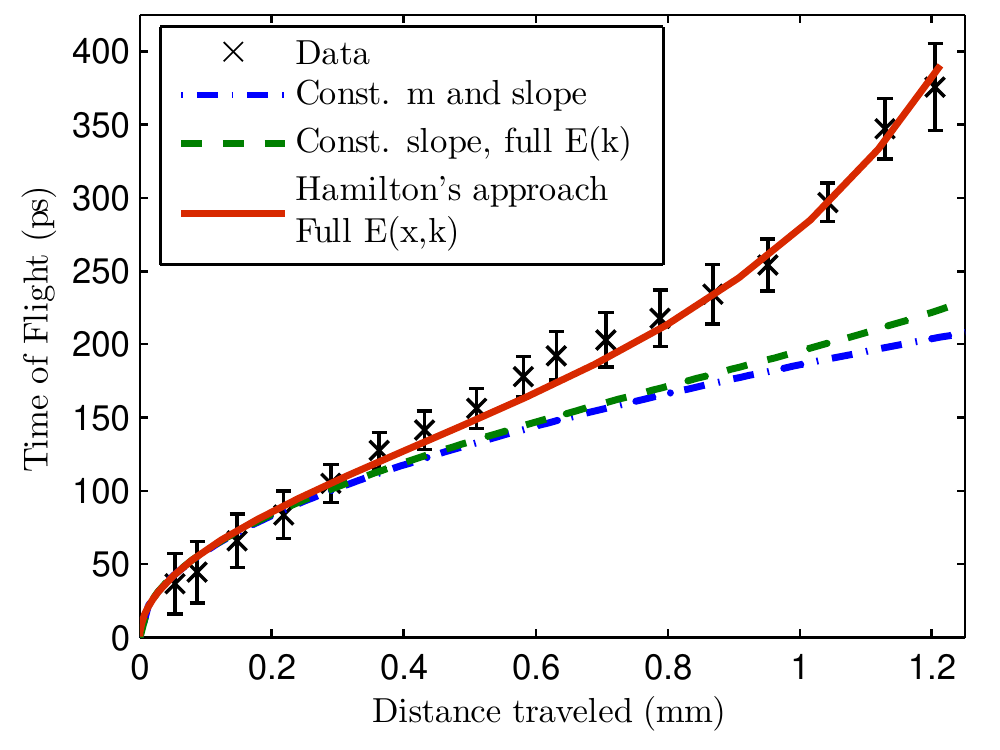}
  \caption{(color online)  Time of flight for polaritons with different initial momenta to reach $k_{\|} =0$ for the same pumping conditions as fig.~\ref{TimeProfiles}.  Black crosses with error bars: the time of flight as determined from the time-resolved data.  Blue solid line:  fit of the data assuming a constant mass and constant gradient of potential energy (i.e., constant force) felt by the polariton.  This model clearly fails to describe the later time arrivals.  Green curve: fit of the data assuming a constant force on the polaritons but allowing for the full dispersion relation $E(k)$ of the polaritons, which has effective mass that changes at higher $k_{\|}$.  Accounting for this changing mass improves the fit only slightly.  Red curve: fit calculated by numerically propagating $x(t)$ and $k(t)$ according to the full semiclassical Hamiltonian of the LP.}
  \label{TimeOfFlight}
 \end{center}
\end{figure}

What is clear from the raw data and from the fits is that the more distant points take longer to be populated with polaritons.  If the motion is ballistic in nature, then we should expect the dynamics to be explained by semiclassical particle dynamics.  In Fig.~\ref{TimeOfFlight} we present the time-of-flight value $TOF=t_0-t_{cool}$ from the fits as a function of the distance from the excitation spot. The cooldown time, $t_{cool}$, is the time for the hot excitations (as observed in Fig.~\ref{EvX_LowDensity}(a)) to fall down into the LP states from which they can begin to propagate.  Note that $t_{cool}$ was determined by fitting the data to theoretical models of propagation, since the data immediately at the excitation point shows unreliable $t_0$ values.  The simplest approach to explain the data is to assume that the polariton mass is constant and the potential gradient felt by the polaritons due to the wafer thickness variation is constant, i.e., that the polaritons feel a constant force.  We can envision the polaritons as starting with an initial momentum uphill and we observe them when they reach $k_{\|}=0$.  This yields the relationship that the time of flight is proportional to $x^{1/2}$, which is shown as the blue dash-dotted curve in Fig.~\ref{TimeOfFlight}.  This works well for short distances, but the data beyond $x=0.4$ mm show a clear upturn which deviates from this simple fit.

To go beyond this simple model, we can recognize that the effective mass approximation breaks down for polaritons at high momentum.  Due to the coupling of the very light mass photon and the heavy mass exciton, the dispersion of the polaritons at high momenta deviates from the effective mass measured at $k_{\|}=0$.  This is particularly true on the photonic side where the region near $k_{\|}=0$ may have a mass on the order of $10^{-4}$ times the electron mass, while larger $k$-values at the same spatial point have a mass on the order of half the electron mass.  By using the known polariton parameters (including the coupling strength between the exciton and photon, the cavity gradient and resonance position), we can relate the distance traveled to the initial energy and therefore the wavevector of the polariton.  If the gradient of the polariton energy is approximately constant, then the force on the polaritons will be constant and the time of flight will depend linearly on the initial wavevector according to $\hbar \partial k/\partial t = F$.  Including the effect of the non-parabolic dispersion relation (green dashed line in Fig.~\ref{TimeOfFlight}) gives a slight upturn in the time of flight at farther distances.  The effect of the increasing mass is to slow the deceleration.  However, this model does not yet fully fit the data.

To accurately fit the data we must take into account the fact that the polariton energy in the strong coupling region near resonance quickly transitions from the rapidly changing photonic energy to the slowly changing exciton energy, and its mass changes by orders of magnitude.  Thus we should not be surprised that naive models assuming constant mass and force will fail.  However, the complicated energy of the polariton $E(x,k)$ prohibits a simple analytical solution to the time of flight as a function of the initial $x$ and $k$.  The most adequate solution to such a problem is directly deriving the equations of motion from the Hamiltonian, $\mathscr{H}(x,k)$, based on the known polariton parameters.

Here we express the time change in $x$ and $k$ via the relationships $\dot{x}=\partial \mathscr{H} / \partial \hbar k$ and $\hbar \dot{k}=- \partial \mathscr{H} / \partial x$. Starting from the initial position and energy (which is assumed to be conserved) we can propagate these values until the final wavevector is zero, which is the emission that we observe in data like that of Fig.~\ref{TimeProfiles}.  Accounting for both the non-trivial dispersion relation and spatial potential yields the red curve in Fig.~\ref{TimeOfFlight}, which follows the data within the uncertainty, even far from the excitation spot.

\section{Estimation of the Polariton Lifetime}

The long-range motion of polaritons in these samples suggests a significantly longer lifetime than has been observed in older samples.  One might look for a direct measurement of the lifetime, but for various reasons this is difficult.  We expect a lifetime on the order of 100 ps, so one might imagine that we can measure the decay of the cavity emission with a streak camera.  However, as discussed above, if we generate the polaritons non-resonantly, this decay will mostly be detecting the thermalization time of hot carriers as they cool and become polaritons. On the other hand, resonant excitation of the polaritons is also problematic. For a measurement of the lifetime we could imagine resonantly exciting a polariton state and measuring the PL emitted from that state. There are several problems with this. First, there will be a large amount of reflected laser light, which can be reduced but not completely eliminated. Second,  the lifetime of this state will mostly be affected by the dynamics of scattering into different polariton states.  Third, with resonant excitation a coherent polariton state is produced which can have superradiant emission.

Another approach would be to measure the linewidth of the cavity photon mode, which will directly give a lower limit to the lifetime.  The spectral resolution of our equipment, however, is not small enough to measure a 100 ps lifetime, which corresponds to a FWHM of less than 7 $\mu$eV.  We measure a line width at the limit of our spectrometer resolution of 0.05 nm (100 $\mu$eV), which implies a lifetime of at least 7 ps.

\subsubsection{Lifetime from time-resolved intensity versus position}

Due to the difficulty of applying these more direct methods of measuring lifetime, we present here our best estimate of the lifetime from two different methods based on understanding the ballistic motion of these long-lived polaritons. Note that the lifetime of the polaritons is inversly proportional to their photonic fraction for photonic detunings. The lifetime is always longer in the excitonic region of the wafer, or in high-$k$ states which have greater excitonic fraction.  We are primarily interested in the intrinsic cavity lifetime, which is half the polariton lifetime at the resonant detuning point where the polaritons have 50\% photon fraction. 

The transport results discussed in the previous sections demonstrate the persistence of polaritons for hundreds of ps following non-resonant excitation---as seen in Fig.~\ref{TimeOfFlight}, the offset time for the arrival of polaritons reaches 400 ps. In addition to measuring the TOF in the above data, we also have measured the overall intensity reaching $k_{\|}=0$ at various positions across the sample.  Each final point corresponds to the number of polaritons that have survived the time of flight.  We expect an exponential decay due to leakage of the photon mode through the mirrors, so the final population should be $n(t_0)=n(0)\exp(-t_0/\tau_i)$ where $t_0$ is the time of flight for that datapoint and $\tau_i$ is the lifetime of that state.   

\begin{figure}[h]
 \begin{center}  
  \includegraphics[width=3.35in]{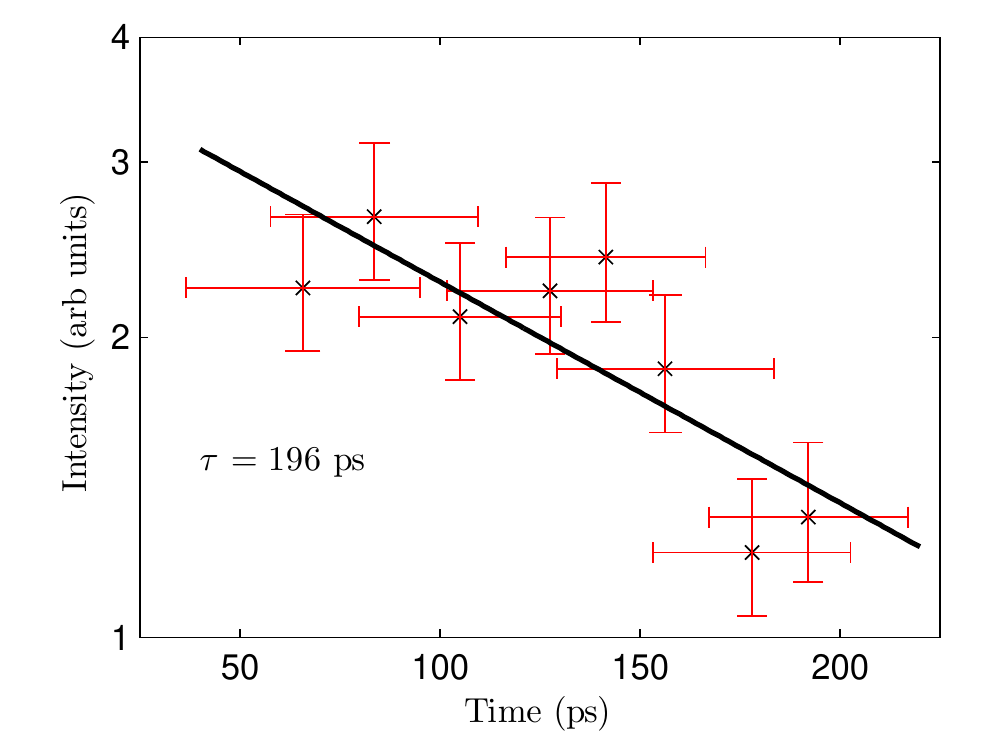}
  \caption{Lifetime of the polaritons based on the normalized intensity versus a time of flight.  The time values for the red crosses (data points) are the time-of-flight data presented in Fig.~\protect\ref{TimeOfFlight}.  The intensity values of these data points are the intensity detected at the point and time of measurement, normalized by the intensity at the same polariton energy taken from $k$-resolved data under the same conditions as Fig.~\ref{EvK_LowDensity}(a) except that the pump spot detuning was the same as the time-resolved conditions. Since the emission at each spatial point corresponds to a single initial $k_{\|}$-state at the pump spot, this ratio gives the loss during the spatial propagation due to radiative emission and other scattering processes.  The solid black line is a fit of a single exponential decay with lifetime of 200 ps.}
  \label{IntensityTOF}
 \end{center}
\end{figure}

Figure~\ref{IntensityTOF} was determined by the following process: 1) the intensity $I(x)$ at $k_{\|} = 0$ was found for a range of distances $x$ from the generation spot. Because of the gradient of polariton energy, each of these positions had a different energy. 2) The initial intensity $I_0(E)$ as a function of energy was found at the generation spot, from $k$-space data such as shown in Fig.~\ref{EvK_LowDensity}(a). The higher energies correspond to higher momenta; these momenta drop to $k_{\|}=0$ as the polaritons travel uphill. 3) The ratio $I(x)/I_0(E(x))$ was plotted as a function of the time-of-flight value $t_0$ found for each value of $x$.  If we assume that the lifetime is approximately constant for polaritons in a certain energy range, then fitting this plot to an exponential decay gives the lifetime. The result of this lifetime fit gives a polariton lifetime of 200 ps, as shown in Fig.~\ref{IntensityTOF}.  We note that this lifetime includes all processes which remove particles from a ballistic path, including scattering from disorder. In addition to showing that the cavity lifetime is long, this measurement also shows that the disorder is very low. 

Of course, the polariton lifetime is not constant, but depends on the energy of the polaritons due to the dependence of the photon fraction on the detuning.  Over the range of energies used in Fig.~\ref{IntensityTOF}, we estimate that the photon fraction changed from about 90\% to 75\%. The fit value for the polariton lifetime of 200 ps therefore represents a cavity lifetime of about 150-180 ps.

%Figure showing I vs time of flight (from x data).

\subsubsection{Lifetime from CW intensity}

\begin{figure}[h]
 \begin{center}  
  \includegraphics[width=3.35in]{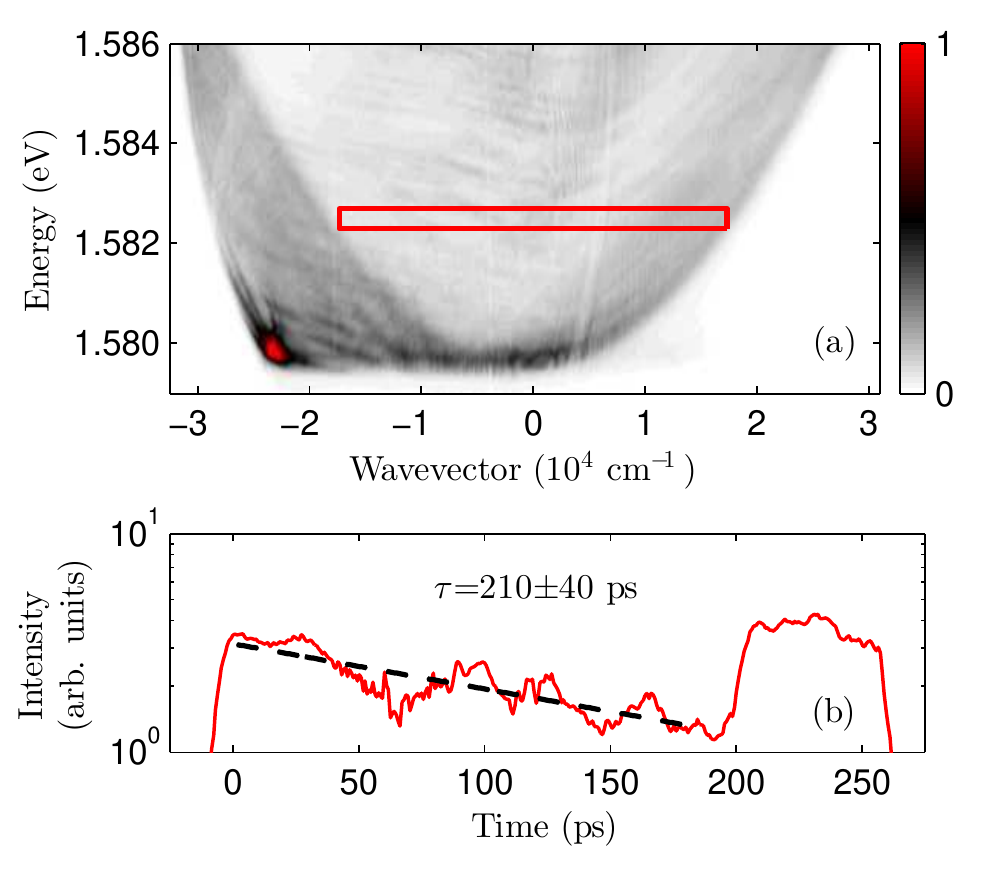}
  \caption{(color online) Lifetime measurement based on the steady state $k$-resolved PL intensity data.  (a) The same data as presented in Fig.~\ref{EvK_LowDensity}(b), with a highlight showing a selected detuning to generate an intensity profile. (b) The intensity profile for the selected detuning. Dashed line: fit to a single exponential decay in time. Note that the polaritons travel uphill and come back down. We therefore restrict the fits to times before the polaritons have returned back to the same place, which corresponds $k_{\|}$ equal but opposite the initial $k_{\|}$.  The polaritons moving downhill from the generation point are ignored due to noise in the data and the fact that they are observed for a short period of time which renders the fits unreliable.  The time calibration in this plot is generated using the $k(t)$ prediction at each energy based on the initial conditions and applying Hamilton's method, as used for the fit of Fig.~\protect\ref{TimeOfFlight}.}
  \label{LifeFromK}
 \end{center}
\end{figure}

An alternative way to measure the lifetime of the polaritons is to track the intensity change in $k$-space.  The fit of the Hamilton's method theory in Fig.~\ref{TimeOfFlight} gives $k(t)$ for each polariton energy. Therefore we can convert $I(k)$ to $I(t)$ for a given energy in data like that of Fig.~\ref{EvK_LowDensity} and extract a lifetime for each polariton energy from a fit to an exponential decay. This is shown in Fig.~\ref{LifeFromK}.  Here the photon fraction ranges from about 95\% at lowest energy to 85\% at highest energy.

%Figure showing I vs calculated time of flight (from k data).

Over the range of detunings with reliable fits, these results show a lifetime of 200 $\pm$ 120 ps in a region where the polariton is mostly photonic.  While we are unable to extract a trend of lifetime vs initial wavevector that clearly matches up with detuning dependent liftime or scattering trends, it is clear that these data support the conclusion that the cavity mode has a lifetime on the order of 100 ps.

\section{Higher Density: Coherent Flow}

As the density is turned up, the polaritons experience a blue shift of their $k_{\|}=0$ energy.  This comes about due to exciton-polariton repulsion and possibly also to some degree due to a shift of the lower polariton branch due to phase space filling, which reduces the oscillator strength that gives the Rabi splitting between the upper and lower polariton branches.  The excitons are produced by the same off-resonant pumping process that generates the polaritons---hot free carriers first form into excitons, and then some fraction of the excitons scatter down into exciton-polariton states.  In many cases the exciton population can be 20 times greater than the polariton population \cite{Hartwell2010}. The exciton population does not move long distances like the polaritons in these samples, however, because the exciton mass is about $10^4$ times larger than the lower polariton mass. The exciton cloud diffuses at most a few microns from the laser excitation spot. This has been used \cite{Cristofolini2013,Wen2013Submission} to create user-controlled potential barriers for polaritons. In many works with short-lifetime polaritons, the exciton cloud is assumed to be everywhere that the polaritons are, and is called the ``exciton reservoir,'' but in our long-lifetime samples, the polaritons can move very far from the exciton cloud.

In the experiments reported here, the polaritons are in an unbounded geometry---they can flow away from the excitation spot in the two-dimensional plane of the microcavity. It is therefore problematic to define Bose-Einstein condensation exactly. In a two-dimensional unbounded system, there is no ``true'' condensation \cite{Holzmann2007,Berman2008}.  Rather, the fraction in low-energy states near the ground state increases rapidly as the density increases, for a constant temperature, until a large fraction of the particles are in states with kinetic energy much less than $k_BT$. This is often called the ``quasicondensate" \cite{David2013} The quasicondensate has many of the properties of a ``true'' condensate but has imperfect phase coherence. 

\begin{figure}[h]
 \begin{center}  
  \includegraphics[width=3.3in]{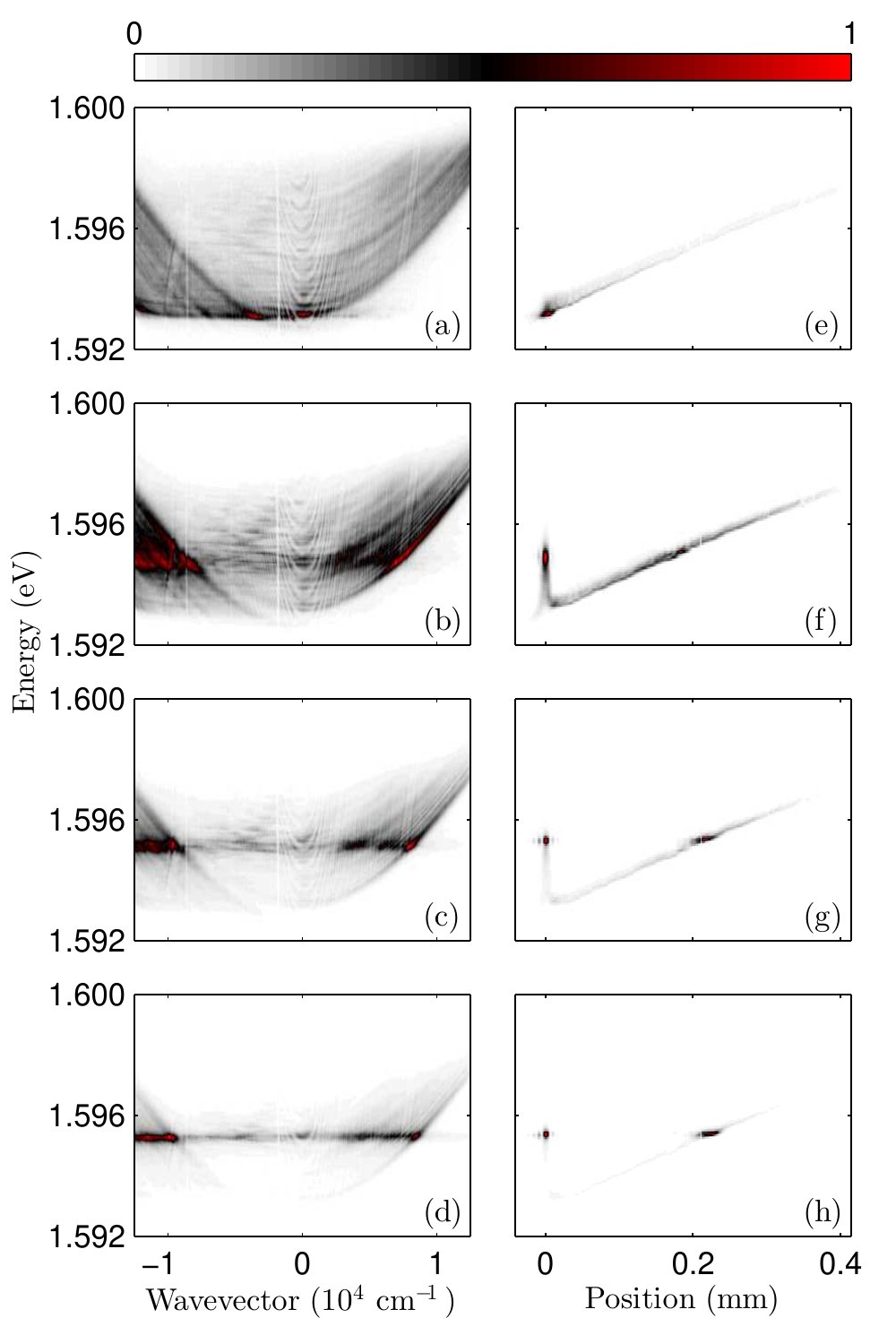}
  \caption{(color online) Figures (a)-(d) show $k$-resolved PL from the polariton population at pump powers of 0.25 mW, 21 mW, 30 mW, and 35 mW, respectively.  These data were collected using a pump laser with wavelength of 705 nm focused to a 15 $\mu$m diameter pump spot where the LP detuning was -8 meV.  Figures (e)-(h) show $k_{\|} \sim 0$ real-space-resolved emission at the same densities.  Note that at the lowest density ((a) and (e)), all of $k$-space is occupied at the emission spot and the polaritons roll uphill and downhill as discussed above.  However, as the pump power increases and renormalization occurs at the pump spot, a larger occupation builds up in the $k_{\|} = 0$ state on top of the potential-energy hill at the pump spot.    The high occupation of a single state is seen as a monoenergetic line in $k$-space and two spots in the low-NA, real-space data, corresponding to the excitation spot and the turnaround point 200 $\mu$m away.  In real space only two spots are observed because the polaritons in between, as well as those traveling downhill, are outside the angle of emission being imaged.}
  \label{EvXK_MidDensity}
 \end{center}
\end{figure}

In the case of a steady-state system with generation, decay, and flow away from the point of creation, the ground state of the system is not localized to just the region where the particles are created. As we have seen in the previous sections, the polaritons can travel ballistically hundreds of microns away from the creation spot. We therefore expect that the ground state will be a state that extends far from the creation point even while having a single energy \cite{Wouters2008}.

Figure \ref{EvXK_MidDensity} shows the real-space and $k$-space energy distribution of the polaritons under similar conditions as Figs. \ref{EvX_LowDensity} and \ref{EvK_LowDensity}, namely off-resonant excitation on the photonic side of the wafer, but with increasing excitation density. Two changes are notable as the density increases. One is that the energy of the polaritons shifts upward. This energy shift corresponds to the shift of the ground state of the polaritons at the point of creation due to their repulsion from the exciton cloud, discussed above. The second notable feature is that the energy distribution of the polaritons changes from a broad range of energies (Cf. Fig.~\ref{EvK_LowDensity}(b)) to a single energy. This is due to the interactions of the polaritons in the excitation region, which allow them to thermalize. Even though they never perfectly thermalize when they are mostly photonic in character, as is the case here, they still have enough interactions to redistribute their energy distribution strongly toward the ground state. As seen in Fig.~\ref{EvXK_MidDensity}(h), they move at the same energy several hundred microns away from the laser excitation spot. Although the polaritons far from the exciton cloud probably have very weak interaction with each other, they still maintain the same energy. This extended, mono-energetic state is the effective ground state of the steady-state system, as discussed above. The two bright spots at $k_{\|} \sim \pm 1\times 10^4$ cm$^{-1}$ in Fig.~\ref{EvXK_MidDensity}(d)  correspond to the velocity which the polaritons have after accelerating away from the exciton cloud, trading all of their potential energy for kinetic energy (cf. Ref.~\onlinecite{Wouters2008} ). The polaritons moving uphill, with initially positive $k_{\|}$, slow down and eventually pass through $k_{\|}=0$, which corresponds to the turnaround point seen in Fig.~\ref{EvXK_MidDensity}(b4). After passing through $k_{\|}=0$, they have turned around and are moving in the opposite direction.

\begin{figure}[h]
 \begin{center}  
  \includegraphics[width=3.3in]{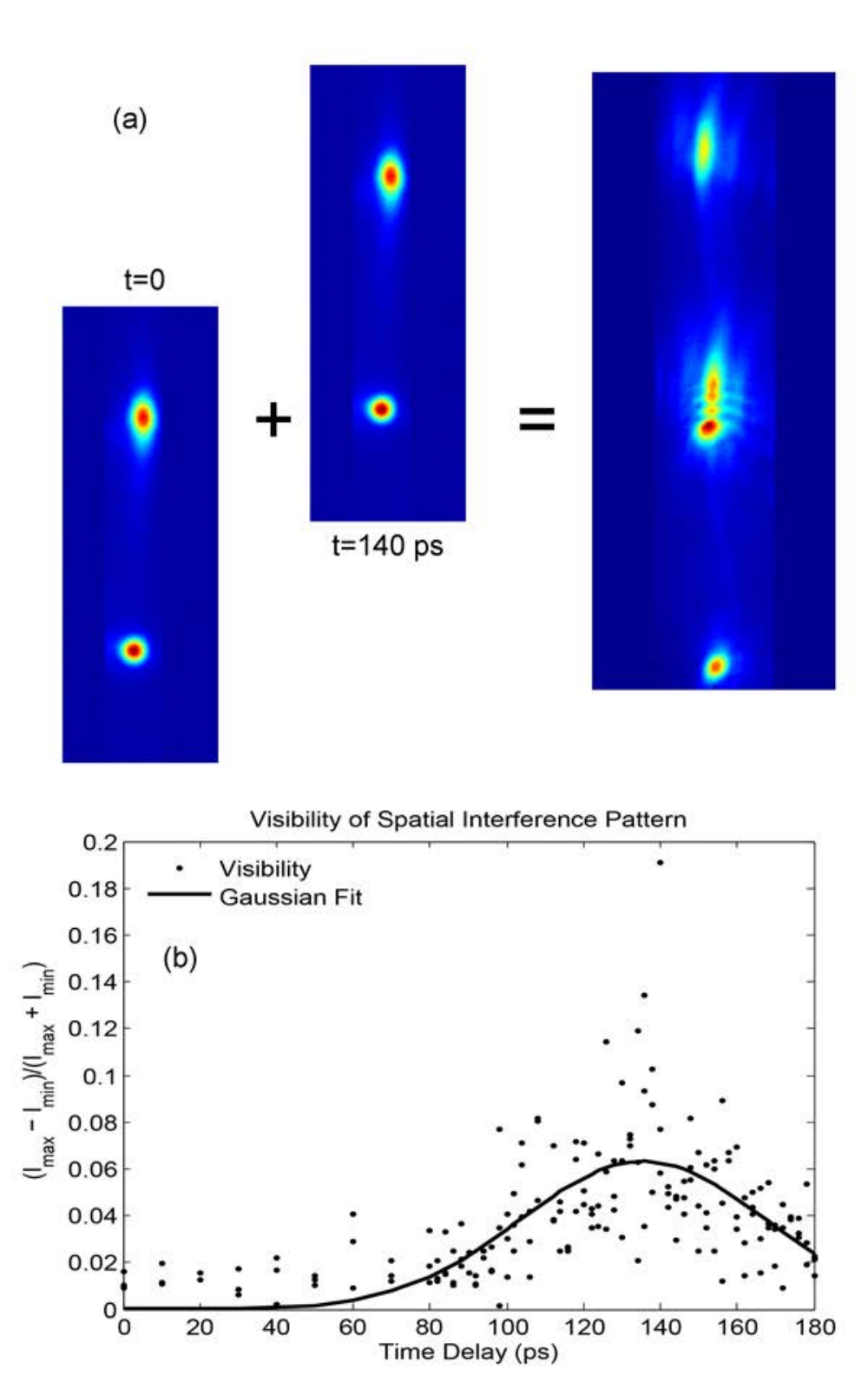}
  \caption{(color online) Interference measurements conducted by overlapping PL from the pump spot with time-delayed PL from the turnaround point in the medium density regime.  These data were collected using a pump laser with wavelength of 705 nm focused to a 25 $\mu$m diameter pump spot where the LP detuning was -4.5 meV. Frame (a) shows the real space luminescence from the individual points and a sample interference pattern.  Frame (b) plots the visibility of the fringes as a function of delay time.  The fact that the greatest visibility is seen at 140 ps makes perfect sense as this is the propagation time for the polaritons to travel 200 $\mu$ m from the pump spot to the turnaround point.  The high scatter and overall low visibility of the fringes is primarily due to the instability in the pump laser, which leads to instability of the blueshift peak on which the polariton quasicondensate is formed and therefore causes the both the condensate energy and turnaround point to fluctuate.}
  \label{InterferenceFig}
 \end{center}
\end{figure}

This monoenergetic quasicondensate acts as coherent wave. One way to see that the state is coherent is to simply note the spectral narrowing, to a peak with width about 0.2 meV.  This width is actually broadened somewhat by the time averaging in our experiments. Fluctuations of the laser power lead to fluctuations of the exciton cloud potential energy height, which determines the polariton ground state energy. Another way to see the degree of coherence is by an interference measurement. Figure~\ref{InterferenceFig}(a) shows the spatial pattern which is the result of interfering the $k_{\|} = 0$ emission from the creation spot with the $k_{\|} = 0$ emission from the turnaround spot 200 $\mu$m away. Figure~\ref{InterferenceFig}(b) shows the fringe contrast as a function of delay time. This shows that the coherence time of the propagating ground state is approximately 40 ps, with an offset given by the propagation time $t_{prop} = 140$ ps from the creation spot to the turnaround spot. We believe that this interference measurement is also somewhat degraded by fluctuations of the pump laser power, which cause not only fluctuation of the energy of the polaritons due to the change of the potential energy of the polaritons due to the exciton cloud density, but also fluctuations of the spatial position of the turnaround point, i.e., the point with $k_{\|}=0$ energy equal to that at the creation point.

This quasi-coherent flow can be easily understood as the propagation of a macroscopically occupied single wavefunction according to the system Hamiltonian.  A simple approximation is to model the evolution of the quasicondensate using a 1D Schr\"odinger equation. While this involves approximations (for example, outflow to the sides will give a shorter effective lifetime), it makes the problem manageable and can recreate the major features of the observed real-space distribution, and allows us to make another constraint on the polariton lifetime.

To model this system we work in the effective mass regime for the lower polariton and model the spatial potential as a linear gradient with a Gaussian peak due to exciton cloud at the excitation spot, as is visible in Fig.~\ref{EvXK_MidDensity}(f) and (g).
This gives the general Gross-Pitaevskii equation
\begin{eqnarray}
i\hbar\frac{\partial \psi}{\partial t} &=& \left(-\frac{\hbar^2}{2m}\frac{\partial^2}{\partial x^2} + U_0e^{-x^2/\sigma^2} + Fx + U|\psi|^2 \right)\psi \nonumber\\
&& - \frac{i}{2\tau}\psi + G(x) ,
\end{eqnarray}
where $U$ is the polariton-polariton interaction potential, $\tau$ is the polariton lifetime, and $G(x)$ is the localized polariton generation term (which can, in general, depend on the local polariton density, since a condensate of polaritons stimulates conversion of excitons into polaritons).  The slope $F$ is measured from the observed polariton gradient at low density, the Gaussian peak height $U_0$ is measured as the condensate emission energy, and the Gaussian peak width $\sigma$ is determined from the pump spot size. The effective mass $m$ can be found from low density $k$-space data (i.e. the curvature of the dispersion seen in Fig.~\ref{EvK_LowDensity}(a)), and we can justify using this effective mass because the mass changes minimally over the narrow energy range of this matter wave. In the low density limit, the polariton-polariton interaction is negligible, and this equation becomes simply a 1D Schr\"odinger equation with generation and decay. 

The eigenstates of the system can be generated for a discretized space by numerically diagonalizing the 1D Schr\"odinger equation.  Once we have a real-space representation of the eigenstates, it is trivial to decompose a matter wavepacket into constituent eigenstates and evolve it.  The finite spatial grid and window leads to quantized states in the downhill direction where there is really a continuum, but artifacts created by this can be minimized if we ensure the space simulated is large enough that the state spacing is small compared to the energy range occupied by the condensate.

Using this prescription, we can evolve the motion of a pulse of matter wave in real space and $k$-space with any lifetime.  We can easily compare the characteristics of different lifetime particles by simply changing lifetime and evolving again. 

\begin{figure}[h]
 \begin{center}  
  \includegraphics[width=3.3in]{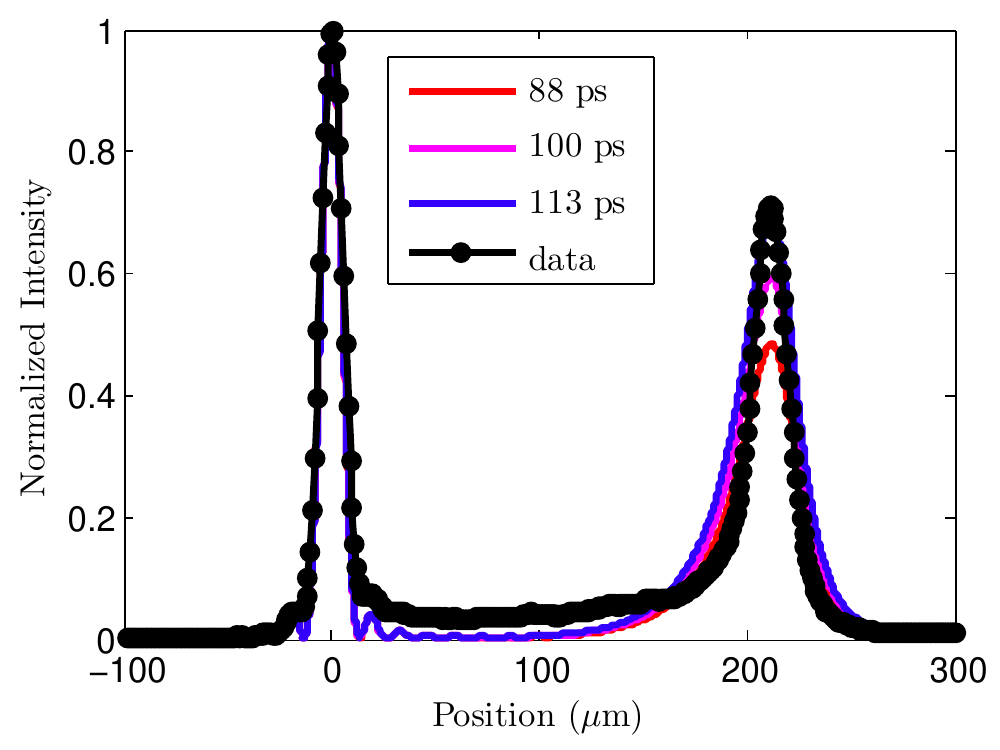}
  \caption{(color online) Comparison of simulated state evolution with observed real space intensity.  The data were collected under the same conditions as Fig.~\ref{InterferenceFig}.  Note that three simulated lifetimes are presented for comparison.}
  \label{Simulation_RealSpace}
 \end{center}
\end{figure}

Simulations with three different lifetimes are presented in Fig.~\ref{Simulation_RealSpace} with comparison to an observed intensity profile with low-NA acceptance.   Comparing the simulation results to $k$-space data also gives good agreement, indicating a good confidence in the simulation parameters such as effective mass and Gaussian peak width.  As seen in Fig.~\ref{Simulation_RealSpace}, changing the lifetime has a strong effect on the relative height of the turn-around intensity peak to that at the generation spot.  A very short lifetime will cause the uphill peak to vanish entirely, as polaritons decay before reaching that point, while a very long lifetime can make the uphill peak intensity comparable to the generation point intensity. The lifetime found here, 113 ps, is an underestimate of the polariton lifetime, because the effective lifetime for this model will be shorter due to outflow of the polaritons in the full 2D system, away from the 1D path considered here.

%Figure showing Mid-density simulation.

Related effects have been seen before with short-lifetime exciton-polaritons. If the laser generation spot is made very small, then there can be separation of the polariton motion and the exciton cloud even if the polariton lifetime is short. Ref. ~\onlinecite{Richard2005,Kasprzak2006b} shows peaks at $\pm k$ which corresponded to acceleration away from the exciton cloud, as here. The group of J. Bloch \cite{Wertz2010} has shown mono-energetic propagation of a quasicondensate in a 1D quantum-confined wire, and Baumberg's group has seen similar behavior \cite{Cristofolini2013} with single laser spots in a 2D unbounded system.

Just as the resonant or non-resonant scheme can affect the polariton condensate formed, Richard et al.\cite{Richard2005} demonstrated that the pump spot can also change the features of the condensate .  It has been observed that a small excitation region can give rise to a condensate at finite $k$ \cite{Wouters2008}.  It is typically the case that polariton condensates form in regions where there is substantial renormalization, since the high carrier density that allows the condensate to form also causes a real blueshift of the polariton. Therefore, it is not surprising that a condensate of small size which is formed on top of a hill will flow outward.

\section{High Density: Trapped Condensate}

While the quasicondensate described in the previous section does not exhibit a sharp threshold, as expected for a  2D system, at higher density we observe a much sharper threshold transition to a trapped condensate with much greater coherence. This has been reported elsewhere \cite{Nelsen2013PRXSubmission}.  In this case the polaritons scatter into a much lower energy state and localize at the energy minimum formed between the renormalized peak and the uphill gradient. 

Although a first glance at the potential energy profile felt by the polaritons would indicate that they are not truly trapped, since the potential energy minimum shown in Fig.~\ref{EvXK_MidDensity} is only in one dimension, there exists a process by which the polariton may in fact self trap, leading to a true 2D confinement. Two terms in the above Gross-Pitaevskii equation should be altered to take into account the interaction of the polaritons and the exciton cloud. First, the generation rate of polaritons can be written as $G(x)(1+\alpha |\psi(x)|^2)$, where $\alpha$ is a parameter, to take into account the fact that high polariton density will stimulate conversion of excitons into polaritons in regions where both exist. Second, the exciton cloud height $U_0$ can be written as $U_0(1-\beta|\psi(x)|^2)$, to take into account the fact that stimulated conversion of excitons will drop the potential energy height felt by the polaritons, since polaritons repel each other more weakly than excitons repel polaritons. The modified Gross-Pitaevskii equation is then
\begin{eqnarray}
i\hbar\frac{\partial \psi}{\partial t} &=& \left(-\frac{\hbar^2}{2m}\frac{\partial^2}{\partial x^2} + U_0(1-\beta |\psi|^2)e^{-x^2/\sigma^2} + Fx \right. \nonumber\\
&&  +U|\psi|^2 \biggr)\psi  - \frac{i}{2\tau}\psi + G(x)(1+\alpha |\psi|^2) .
\end{eqnarray}
This highly nonlinear equation can have self-trapping solutions near the exciton cloud. 

When multiple laser spots are used, an externally generated trapping potential can be created. Then even when the polaritons are generated in a region of the wafer where they are more exciton-like, they can undergo Bose condensation to a trapped state very much like the one reported in Ref.~\onlinecite{Nelsen2013PRXSubmission}. The increased lifetime of the polaritons allow for better thermalization of the polariton gas and truly equilibrium condensate theory to apply \cite{Wen2013Submission}.  

\section{Conclusions}

With increased reflectivity on the mirrors in these new high Q-factor microcavity structures, the polaritons demonstrate qualitatively different phenomena.  Even in the low density regime we observe clear signs of polaritons propagating much farther than previous samples with or without 1D waveguide structures which promote  long-range motion.  At higher density we observe long-range, monoenergetic outflow which can be interpreted as a quasicondensate due to the Bose statistics of the interacting polaritons.  The outflow from this condensate carries its coherence over a long distance.

These phenomena are a direct result of the increased lifetime of the polariton, and they also give us indirect ways to estimate the polariton lifetime.  More direct methods of measuring the lifetime are diffucult due to the very narrow linewidth of the cavity photon and high reflectivity of the cavity.  However, by looking at the decay of the polaritons with distance in real space and $k$-space we can have estimated the lifetime of the polariton to be greater than 100 ps, about an order of magnitude longer than previous samples.

\section{Acknowledgments}

The work at the University of Pittsburgh was supported by the National Science Foundation under Grant No. DMR-1104383. The work at Princeton University was partially funded by the Gordon and Betty Moore Foundation as well as the National Science Foundation MRSEC Program through the Princeton Center for Complex Materials (DMR-0819860).

\section{Appendix: Gaussian-exponential decay convolution and fitting}

The form of the time-resolved polariton PL can be understood best as the result of hot excitations relaxing into the polariton states.  The rise time indicates a multiple-path relaxation from the hot excitations to the LP state, so the complicated dynamics become difficult to model.  Since we cannot measure the intermediate or high energy populations, the uncertainty in the parameters governing the relaxation becomes very large.  Because of this, we use simple functions to parameterize the data.

A convolution of a Gaussian with an exponential decay was chosen as a natural function to fit the observed time-resolved PL data with a minimum number of fit parameters.  The data clearly exhibits a long decay time which suggests fitting the data with an exponential decay, and the rise time fits a Gaussian broadening  reasonably well; the broadening can be understood as due to the multiple paths for polariton generation from the initial incoherent hot carriers created by the pump laser.  The central time of the Gaussian peak gives a convenient parameter to measure the arrival time of the polariton population.  Including the overall intensity of the data and background, this means that each curve is fit with 5 parameters.

The Gaussian-exponential convolution (GEC) is calculated according to 
\begin{equation}
n(t) =\int_{0}^{\infty}\Bigl(\Bigl[\frac{A}{\sigma\sqrt{2\pi}}e^{\bigl(\frac{-(t-x-t_0)^2}{2\sigma^2}\bigr)}\Bigr] 
\Bigl[\frac{1}{\tau}e^{\bigl(-x/\tau\bigr)}\Bigr]\Bigr) dx .
\label{GEconvStart}
\end{equation}

The five parameters of the model are $\sigma$, the broadening of the Gaussian; $t_0$, the peak time of the unconvoved Gaussian; $\tau$, the exponential lifetime; $A$, the time-integrated intensity, and ultimately a possible background.  Performing the convolution leads to the form 
\begin{equation}
n(t) = \frac{A}{2\tau}e^{\bigl(\frac{\sigma^2-2 t \tau+2 t_0 \tau}{2 \tau^2}\bigr)} \erfc\bigl(\frac{\sigma^2+t_0 \tau-t \tau}{\sqrt{2} \sigma \tau}\bigr)
\label{GEconvEnd}
\end{equation}
where erfc$(t)$ is the complimentary error function.

Since the GEC model is not derived from a theoretical basis of the relaxation of excitations to the LP states, it is dangerous to interpret too much from the parameters of the fit.  For example, the decay time $\tau$ is not simply the lifetime of the LP population; it includes the effect of the mean lifetime of the reservoir particles to scatter into the LP state.  If the excited states, that is, hot free carriers and excitons, take a long time to relax but have no other means to decay quickly, then it is possible to measure a long lifetime for this decay parameter even if the final polariton decay process is fast \cite{Nardin2009}.  However, we note that the rise time to populate the polariton states is on the order of 80 ps, which is not substantially longer than the decay time itself, and the range of decay times measured from these fits are on the same order as the other lifetime estimates, so these values are still in agreement with our assessment that the polaritons themselves have a lifetime on the order of 100-200 ps.

While several parameters of this fit do not directly give information about the polariton dynamics, the $t_0$ parameter is useful and indicative of the time of arrival of the polaritons at the location being observed.  Other methods of assigning this time, such as the peak of the time-intensity tail, the 10\% and 50\% turn on times were investigated as well.  While all of these data clearly have different offsets, the overall trends fall within their respective uncertainties.  These assessments were included in assigning the uncertainty of the time-of-flight data, for example in Figs. \ref{TimeOfFlight} and \ref{IntensityTOF}.

%%%%%%%%%%%%%%%%%%%%%%%%%

\bibliographystyle{apsrev4-1}
%\bibliography{library}
\bibliography{BallisticMotion_Submission.bbl}

\end{document}